\documentclass[fleqn,usenatbib]{mnras}

\usepackage[T1]{fontenc}

\usepackage{amsmath}

\usepackage{newtxtext,newtxmath}

\usepackage{graphicx}
\usepackage{xcolor}

\usepackage{bm}
\usepackage{placeins}

\usepackage{hyperref}
\hypersetup{colorlinks=true,linkcolor=blue,citecolor=blue,urlcolor=blue}
\usepackage{orcidlink}

\hypersetup{colorlinks=true,linkcolor=blue,citecolor=blue,urlcolor=blue}
\title[Quenching in the Green Valley at Low Redshift]{Quenching pathways in the green valley at low redshift: confronting SDSS AGN hosts with IllustrisTNG and EAGLE}

\author[G. Gawade]{
Gaurav Gawade \orcidlink{0009-0002-2177-9941}$^{1}$\thanks{E-mail: gauravgawade@proton.me}\\
$^{1}$Department of Physics, St. Xavier’s College, 5, Mahapalika Marg, Mumbai 400001, India\\
}

\date{Accepted XXX. Received YYY; in original form ZZZ}
\pubyear{2025}

\begin{document}
\label{firstpage}
\pagerange{\pageref{firstpage}--\pageref{lastpage}}
\maketitle

\begin{abstract}
We compare low-redshift ($z<0.1$) BPT-selected \emph{pure} optical AGN hosts in SDSS DR7 to colour-selected ``green-valley'' analogue central galaxies in IllustrisTNG100 and EAGLE Ref-L0100N1504. To reduce cross-dataset systematics, we define the green valley internally using $(g-r)$ percentiles: for galaxies with $\log_{10}(M_\star/\mathrm{M_\odot})>10$, we select the 75th--95th percentiles (SDSS observed-frame fibre colours; simulations rest-frame synthetic colours within 30~kpc). SDSS hosts are linked to the MPA--JHU catalogue for stellar masses and aperture-corrected total SFRs. TNG green-valley centrals are almost entirely quenched, with a sharp pile-up at the imposed SFR floor and median $\log_{10}\mathrm{sSFR}\simeq -14.85$ ($\sim 3.5$~dex below SDSS). EAGLE instead produces a broad, continuous distribution with median $\log_{10}\mathrm{sSFR}\simeq -11.71$ and substantial overlap with SDSS, robust to varying the lower percentile between 60 and 90. At fixed mass, TNG yields higher green-valley occupancy fractions (reaching $\gtrsim 60$~per~cent near $M_\star\sim 10^{11}\,\mathrm{M_\odot}$) than EAGLE (20--40~per~cent). A simple forward model of nebular line ratios places EAGLE analogues across the star-forming and composite loci in the BPT plane, while TNG analogues concentrate in a LINER-like, low-sSFR regime. We infer that TNG's kinetic mode drives an efficient, near-binary shutdown of star formation, whereas EAGLE's stochastic thermal feedback supports a slower decline more consistent with local AGN hosts. All catalogues and analysis scripts are publicly released.
\end{abstract}


\section{Introduction}
\label{sec:intro}

Feedback from actively accreting supermassive black holes is now widely recognised as a key ingredient in models of galaxy formation. In cosmological simulations, AGN feedback is required to suppress star formation in massive haloes, reproduce the observed colour bimodality, and regulate the high-mass end of the stellar mass function \citep[e.g.][]{Somerville2015,Croton2016}. Observationally, the mass of the central black hole (or proxy quantities such as bulge mass and central stellar velocity dispersion) correlates tightly with galaxy quiescence at $z\sim 0$, more strongly than halo mass or environment \citep[e.g.][]{Bluck2020,Terrazas2020,Piotrowska2022}. These trends strongly suggest a causal role for AGN feedback, acting alongside other quenching channels such as virial shock heating, morphological stabilisation, and environmental processes.

Despite this broad consensus, the \emph{timescale} and \emph{mode} of AGN-regulated quenching remain debated. Studies of AGN host galaxies have argued for both rapid, ``catastrophic'' shutdowns of star formation and slower, Gyr-scale declines, particularly in disc-dominated systems \citep[e.g.][]{Schawinski2014,Harrison2017}. The so-called ``green valley'' in colour--mass space---between the blue star-forming cloud and the red quiescent sequence---provides an especially sensitive laboratory for probing these pathways. The incidence and properties of AGN hosts in the green valley constrain how often, and for how long, galaxies linger in the transitional phase.

Large-volume hydrodynamical simulations provide a self-consistent playground for exploring quenching physics. The IllustrisTNG (TNG100-1) \citep{Pillepich2018,Weinberger2017,Nelson2018} and EAGLE \citep{Schaye2015,Crain2015} suites both implement sub-grid prescriptions for black-hole accretion and AGN feedback tuned to match global galaxy statistics. TNG employs a dual-mode scheme in which a high-accretion thermal ``quasar'' mode transitions at low Eddington ratios to a kinetic wind mode that deposits momentum and energy into the circumgalactic medium \citep{Weinberger2017,Zinger2020}. EAGLE adopts a single-mode, stochastic thermal feedback model in which accretion energy is accumulated and then injected in discrete heating events \citep{Schaye2015,McAlpine2017}. While both frameworks produce quenched massive galaxies, they differ markedly in how strongly and how abruptly they suppress star formation in massive centrals, with lingering tensions relative to low-redshift data \citep[e.g.][]{Bluck2016,Donnari2019,Piotrowska2022}.

Connecting such simulations to observations is non-trivial. Optical AGN in SDSS are typically identified via the Baldwin--Phillips--Terlevich (BPT) diagram \citep{Baldwin1981}, using fibre-based emission-line ratios and aperture-corrected star-formation rates drawn from value-added catalogues \citep[e.g.][]{Brinchmann2004}. Demarcation curves such as those of \citet{Kauffmann2003} and \citet{Kewley2001} are then used to separate star-forming, composite, and AGN-dominated systems. Simulations, by contrast, provide integrated SFRs and dust-dependent or dust-free synthetic photometry, but no direct line ratios. Simple comparisons based on fixed absolute colour or sSFR cuts are therefore susceptible to calibration offsets and dust treatment differences, and may obscure physically meaningful discrepancies in quenching pathways.

In this work we address these challenges with a deliberately conservative and reproducible construction. We define green-valley galaxies in each dataset (SDSS, TNG100, and EAGLE) using \emph{internal} colour percentiles: among massive galaxies with $\log_{10}(M_\star/\mathrm{M_\odot})>10$, the green valley is taken to be the 75th--95th percentile range of the $(g-r)$ colour distribution (observed-frame fibre $(g-r)$ in SDSS; rest-frame synthetic $(g-r)$ within 30~kpc in the simulations). This rank-ordered definition strongly reduces sensitivity to differences in dust modelling, stellar-population synthesis, and photometric aperture between datasets, while still isolating the reddest $\sim 20$~per~cent of massive galaxies short of the extreme red sequence. Because we do not impose an explicit AGN-luminosity or accretion-rate selection in the simulations, we interpret the simulated samples as colour-selected analogues and focus on the distribution of host-galaxy properties through the transitional regime.

We then compare these simulation-based analogues to BPT-selected pure AGN hosts in SDSS DR7 at $z<0.1$, linked to the MPA--JHU value-added catalogue for stellar masses and total SFRs. In the simulations we select central galaxies above the same stellar-mass threshold and measure stellar masses and SFRs within 30~kpc apertures, following common practice in the TNG and EAGLE literature. To relate the simulated galaxies back to the BPT diagram we forward-model approximate emission-line ratios for the simulations using simple, mass-based metallicity calibrations; this mapping is intended to locate simulated populations in BPT space rather than perform detailed nebular modelling.

This paper addresses three main questions:
\begin{enumerate}
    \item Do current cosmological simulations reproduce the stellar masses, colours, and sSFR distributions of low-redshift BPT-selected AGN hosts observed in SDSS when simulated galaxies are selected as colour-defined green-valley analogues?
    \item How do the different AGN feedback implementations---kinetic in TNG versus thermal in EAGLE---manifest in the demographics of massive transitional galaxies?
    \item Are any discrepancies between simulations and observations robust to changes in the precise green-valley definition and to sampling variance?
\end{enumerate}

Using a single, openly released Python analysis pipeline, we find that the kinetic AGN feedback implementation in TNG leads to an almost binary quenching behaviour, with green-valley analogues that are already fully quenched by the time they meet the colour criterion. EAGLE, in contrast, supports a broad continuum of sSFRs in the green valley that is much closer to the SDSS AGN host population. In other words, previous studies have largely shown that some simulations can reach the correct \emph{destination} in terms of the global quenched fraction, whereas here we focus explicitly on the \emph{journey}: the distribution of specific star-formation rates along the green-valley pathway. The overall analysis workflow is summarised in Fig.~\ref{fig:flowchart}. Throughout, we emphasise simple, transparent statistics and calibrations to facilitate future extensions and comparisons.

\section{Data}
\label{sec:data}

\subsection{SDSS DR7 AGN hosts and control sample}
\label{subsec:data_sdss}

Our observational sample is drawn from the Sloan Digital Sky Survey Data Release~7 (SDSS DR7) legacy spectroscopic galaxy catalogue \citep{Abazajian2009}. We restrict to the main galaxy sample at redshifts $z<0.1$ and link to the MPA--JHU value-added products \citep{Brinchmann2004} for stellar masses, fibre and aperture-corrected total star-formation rates (SFRs), and emission-line fluxes measured within the $3^{\prime\prime}$ fibre. In the specific DR7 MPA--JHU products used here, the emission-line table does not include \texttt{MJD}; we therefore verify that \texttt{PLATEID} and \texttt{FIBERID} match entry-by-entry between the base SDSS catalogue and the emission-line table and then combine them via strict row-index concatenation to avoid spurious many-to-many merges. Stellar-mass and total-SFR tables are taken from the corresponding MPA--JHU DR7 products in the same ordering.

We require signal-to-noise ratio (S/N) $>3$ in each of H$\alpha$, H$\beta$, [N\,\textsc{ii}]~$\lambda6584$, and [O\,\textsc{iii}]~$\lambda5007$, and discard objects with missing stellar masses or SFRs. These cuts yield 250\,049 galaxies with robust emission-line measurements at $z<0.1$. Because these emission-line S/N requirements bias against weak-lined galaxies (and therefore against the most strongly quenched systems), our SDSS samples should be interpreted as emission-line-detected galaxies suitable for BPT classification rather than a complete census of all low-sSFR objects. Fibre colours are defined as $(g-r)=\mathrm{PLUGMAG}_g-\mathrm{PLUGMAG}_r$, where \texttt{PLUGMAG} values are observed-frame fibre magnitudes; at $z<0.1$, $k$-corrections are small and fibre colours effectively trace the nuclear stellar populations hosting the AGN.

AGN classification follows the classical BPT diagram \citep{Baldwin1981}. Using emission-line ratios from the MPA--JHU measurements, we compute
\begin{equation}
  x \equiv \log_{10}\!\left([\mathrm{N\,II}]\lambda6584/\mathrm{H\alpha}\right), \qquad
  y \equiv \log_{10}\!\left([\mathrm{O\,III}]\lambda5007/\mathrm{H\beta}\right),
\end{equation}
and apply the \citet{Kauffmann2003} and \citet{Kewley2001} demarcation curves (see Appendix~\ref{sec:appendix_bpt} and Fig.~\ref{fig:bpt_sdss}). Galaxies above the Kewley curve are classified as ``pure'' AGN, those between the Kewley and Kauffmann curves as composite systems, and those below the Kauffmann curve as star-forming.

For the purposes of this paper we define the \emph{AGN host} sample as all galaxies lying above the \citet{Kewley2001} line. This conservative choice ensures high purity at the expense of some completeness, and matches the selection implemented in our analysis pipeline. Our final pure-AGN host sample used in the analysis contains $N_{\rm AGN}=19\,039$ galaxies.

Stellar masses $M_\star$ and total SFRs are taken from the MPA--JHU catalogue. We define specific SFRs as
\begin{equation}
  \log_{10}\mathrm{sSFR}
  = \log_{10}\!\left(\mathrm{SFR}_{\rm tot} / \mathrm{M_\odot\,yr^{-1}}\right)
    - \log_{10}\!\left(M_\star / \mathrm{M_\odot}\right),
\end{equation}
adding a small numerical floor of $10^{-4}\,\mathrm{M_\odot\,yr^{-1}}$ to $\mathrm{SFR}_{\rm tot}$ before taking logarithms to avoid undefined values for galaxies with formally zero SFR. We adopt the same numerical floor consistently for the simulation samples to ensure a uniform definition of $\log_{10}\mathrm{sSFR}$ across datasets.

For comparison we construct a non-AGN control sample of purely star-forming galaxies lying below the \citet{Kauffmann2003} curve and satisfying the same S/N and redshift cuts. To minimise biases in redshift and apparent magnitude, we match this control sample to the AGN hosts in the two-dimensional plane of redshift and $r$-band fibre magnitude. Specifically, we bin both populations in a $50\times 50$ grid in $(z,\mathrm{PLUGMAG}_r)$ and, in each occupied bin, randomly select star-forming galaxies to match the number of AGN hosts, using a fixed random seed for reproducibility. This yields a control sample of $N_{\rm ctrl}=19\,039$ galaxies with a nearly identical redshift and fibre-magnitude distribution to the AGN hosts.

Throughout the main analysis we restrict both AGN and control samples to $\log_{10}(M_\star/\mathrm{M_\odot})>10$ when comparing to simulations, ensuring completeness and a common stellar-mass range.

\subsection{IllustrisTNG-100}
\label{subsec:data_tng}

For the theoretical comparison we use the TNG100-1 run of the IllustrisTNG suite \citep{Pillepich2018,Weinberger2017,Nelson2018}, a $(\approx 110.7~\mathrm{Mpc})^3$ volume evolved with the moving-mesh code \textsc{arepo}. We analyse the $z=0$ snapshot (snapshot~99). Haloes and galaxies are identified with the \textsc{subfind} algorithm; we focus on central galaxies, defined as the most massive subhalo in each friends-of-friends group, using the \texttt{GroupFirstSub} flag to identify centrals.

Stellar masses are taken from the public 30~kpc aperture catalogues and converted to physical units using $h=0.6774$. We define $\log_{10}(M_\star/\mathrm{M_\odot})$ from the 30~kpc stellar mass and impose a floor at $10^9\,\mathrm{M_\odot}$ in the initial cleaning, tightening to $10^{10}\,\mathrm{M_\odot}$ for the massive-galaxy analyses. Instantaneous SFRs are obtained by summing the star-formation rates of gas cells within the same 30~kpc aperture. As for SDSS, we define
\begin{equation}
  \log_{10}\mathrm{sSFR}
  = \log_{10}\!\left[\mathrm{SFR}_{30\mathrm{kpc}} + 10^{-4}\right]
    - \log_{10}\!\left(M_\star / \mathrm{M_\odot}\right).
\end{equation}

For photometry we use the rest-frame SDSS-like $g$- and $r$-band magnitudes from the public TNG stellar-photometry catalogue, adopting the Charlot \& Fall dust-attenuated set (\texttt{cf00dust}) within a 30~kpc aperture. Colours are defined as $(g-r) = g_{\rm cf00dust} - r_{\rm cf00dust}$.

After applying the centrals selection and basic cleaning, the TNG100 massive-galaxy sample with $\log_{10}(M_\star/\mathrm{M_\odot})>10$ contains 4\,907 centrals, from which we define the TNG green-valley analogues as described in Section~\ref{sec:methods_gv}.

\subsection{EAGLE Ref-L0100N1504}
\label{subsec:data_eagle}

We also use the EAGLE Ref-L0100N1504 run \citep{Schaye2015,Crain2015}, a $(100~\mathrm{Mpc})^3$ volume evolved with a modified version of \textsc{gadget-3}. We analyse snapshot~28, corresponding to $z=0$, providing a local-Universe comparison consistent with the SDSS $z<0.1$ selection. Galaxies are identified with \textsc{subfind}; we restrict attention to central galaxies (\texttt{SubGroupNumber} $=0$) as in TNG.

Stellar masses are taken from the public catalogues, converted to physical units using $h=0.6777$ where required and measured within a 30~kpc aperture. Instantaneous SFRs are computed by summing the SFRs of star-forming gas particles within the same aperture. Specific SFRs are then defined analogously to TNG, with the same $10^{-4}\,\mathrm{M_\odot\,yr^{-1}}$ floor on SFR.

For photometry we use the intrinsic (dust-free) rest-frame $g$ and $r$ magnitudes (\texttt{g\_nodust}, \texttt{r\_nodust}) measured within 30~kpc and define $(g-r) = g_{\rm nodust} - r_{\rm nodust}$. We have verified that applying the empirical dust model of \citet{Trayford2017} to derive attenuated magnitudes does not alter the rank ordering of $g-r$ colours among massive centrals and therefore does not affect the percentile-based green-valley selection used in this work.

After restricting to snapshot~28, centrals, and $\log_{10}(M_\star/\mathrm{M_\odot})>10$, the EAGLE massive-galaxy sample contains 2\,258 centrals. The corresponding EAGLE green-valley analogues are defined in Section~\ref{sec:methods_gv}.

\section{Methods}
\label{sec:methods}

\subsection{Green-valley definition}
\label{sec:methods_gv}

Directly comparing absolute optical colours between SDSS and simulations is complicated by differences in stellar population synthesis, dust treatment, fibre versus aperture photometry, and $k$-corrections. To minimise such systematics we adopt a percentile-based, internal definition of the green valley in each dataset.

For SDSS, TNG100, and EAGLE separately we:
\begin{enumerate}
    \item restrict to a mass-limited parent sample of galaxies with $\log_{10}(M_\star/\mathrm{M_\odot})>10$ (centrals only in the simulations);
    \item construct the distribution of $(g-r)$ colours for this massive parent sample (observed-frame fibre colours in SDSS; rest-frame synthetic colours in the simulations);
    \item compute the 75th and 95th percentiles of the $(g-r)$ distribution; and
    \item define green-valley galaxies as those with colours between these two percentile values.
\end{enumerate}
By construction, the green valley comprises galaxies in the 75th--95th percentile range of the massive-galaxy colour distribution in each dataset, i.e.\ the reddest $\sim 20$~per~cent short of the extreme red tail.

In TNG100, this procedure yields a relatively narrow window at $g-r \approx 0.371$--0.396; in EAGLE, the intrinsically redder and broader colour distribution places the green valley at $g-r\approx 0.689$--0.783 (Fig.~\ref{fig:gv_definition}). The SDSS green-valley window is defined analogously using fibre colours. These simulation-specific thresholds are used consistently when selecting green-valley analogues and when computing occupancy fractions.

To test robustness, we repeat the analysis with a family of alternative green-valley definitions in which the upper percentile is fixed at the 95th percentile while the lower percentile is varied between the 60th and 90th percentiles in steps of 5~per~cent. For each choice we recompute the median sSFR of the resulting green-valley samples in SDSS, TNG, and EAGLE (Fig.~\ref{fig:percentile_sweep}).

\subsection{Derived quantities and statistical methods}
\label{subsec:methods_stats}

All stellar masses and SFRs are expressed in solar units; we work with $\log_{10}M_\star$ and $\log_{10}\mathrm{sSFR}$ throughout. Kernel density estimates (KDEs) for one-dimensional distributions are computed using \textsc{seaborn}, adopting Scott's rule for the bandwidth and using consistent smoothing settings (e.g.\ the same \texttt{bw\_adjust}) across samples within a given figure.

Median relations (e.g.\ $\log_{10}\mathrm{sSFR}$ as a function of $\log_{10}M_\star$) are computed in fixed-width stellar-mass bins, typically $\Delta\log_{10}M_\star=0.2$, requiring at least $10$ galaxies per bin; the resulting medians are shown with interquartile ranges where relevant. All quoted medians and their uncertainties are estimated via bootstrap resampling.

Specifically, for any quantity $x$ we draw $N_{\rm boot}=5000$ bootstrap resamples of the sample, compute the median in each resample, and report the central value and error bars as
\begin{equation}
  \tilde{x} = \mathrm{median}(x), \qquad
  \Delta x_+ = x_{84} - \tilde{x}, \qquad
  \Delta x_- = \tilde{x} - x_{16},
\end{equation}
where $x_{16}$ and $x_{84}$ are the 16th and 84th percentiles of the bootstrap median distribution. This procedure is implemented by the global helper \texttt{get\_bootstrap\_stats} defined in the analysis notebook.\footnote{The full notebook and helper functions are archived on Zenodo together with this manuscript; see the Data Availability section.}

To quantify differences between distributions we use the two-sample Kolmogorov--Smirnov (K--S) test, reporting the K--S statistic $D$ and its associated $p$-value for the null hypothesis that the two samples are drawn from the same parent distribution. Because the SDSS samples are much larger than the simulation samples, the raw $p$-values are often formally vanishing; to assess the magnitude of the discrepancies we therefore complement these tests with bootstrap experiments in which we repeatedly resample SDSS to match the simulation sample sizes and recompute $D$ (Fig.~\ref{fig:bootstrap_ks}).

Green-valley occupancy fractions (Fig.~\ref{fig:gv_occupancy}) are defined in fixed stellar-mass bins as
\begin{equation}
    f_{\rm GV}(M_\star) = \frac{N_{\rm GV}(M_\star)}{N_{\rm tot}(M_\star)},
\end{equation}
where $N_{\rm GV}$ is the number of galaxies in the green-valley colour window and $N_{\rm tot}$ is the number of galaxies in the parent sample (AGN hosts, control galaxies, or all centrals in a simulation). We estimate binomial uncertainties as
\begin{equation}
    \sigma_f = \sqrt{\frac{f_{\rm GV}(1-f_{\rm GV})}{N_{\rm tot}}},
\end{equation}
and omit bins with fewer than 10 galaxies.

\subsection{Forward-modelling of BPT line ratios}
\label{subsec:methods_bpt}

To provide qualitative context for how simulated galaxies populate classical optical-emission-line phase space, we construct approximate emission-line ratios for TNG and EAGLE galaxies and place them in the BPT plane. We do not perform full photoionisation or radiative-transfer calculations; instead, we build a simple empirical mapping from stellar mass to metallicity and then to expected nebular line ratios, capturing the first-order structure of the star-forming sequence.

We adopt a linear mass--metallicity relation, chosen to approximate local observations,
\begin{equation}
  12 + \log(\mathrm{O/H}) = 8.69 + 0.30\,[\log_{10}(M_\star/\mathrm{M_\odot})-10],
\end{equation}
and convert metallicity to approximate line ratios using analytic fits to the observed star-forming locus:
\begin{align}
  \log_{10}\frac{[\mathrm{N\,II}]}{\mathrm{H\alpha}} &\approx
    \frac{12 + \log(\mathrm{O/H}) - 8.90}{0.57}, \\
  \log_{10}\frac{[\mathrm{O\,III}]}{\mathrm{H\beta}} &\approx
    -0.25\,[12 + \log(\mathrm{O/H}) - 8.70] + 0.5.
\end{align}
We add Gaussian scatter to both coordinates (with dispersions chosen to reproduce the width of the SDSS star-forming sequence) and discard points outside the standard plotting window. The resulting mock TNG and EAGLE BPT points are colour-coded by their sSFR and compared to the SDSS distribution (Fig.~\ref{fig:bpt_mock}). We also report the fraction of mock points lying above the \citet{Kewley2001} line as an illustrative measure of occupancy of the formal AGN region under this mapping; in practice this fraction is computed only over the standard domain of the Kewley curve ($\log([\mathrm{N\,II}]/\mathrm{H\alpha})<0.47$) and depends on the adopted scatter in the toy mapping. Because explicit AGN ionisation is not modelled, this should not be interpreted as a prediction of the true AGN fraction.

This forward-modelling is used only for qualitative, phase-space comparisons and to estimate the relative occupancy of different BPT regions. It does not enter the main green-valley selection or any of the quantitative sSFR statistics. We emphasise that this forward model is a simplified metallicity-based mapping intended for phase-space visualisation; it does not account for complex ionisation-parameter variations or shock heating present in real AGN and LINER-like systems.

\subsection{Reproducibility and workflow}
\label{subsec:methods_repro}

\begin{figure*}
\centering
\includegraphics[width=0.85\textwidth]{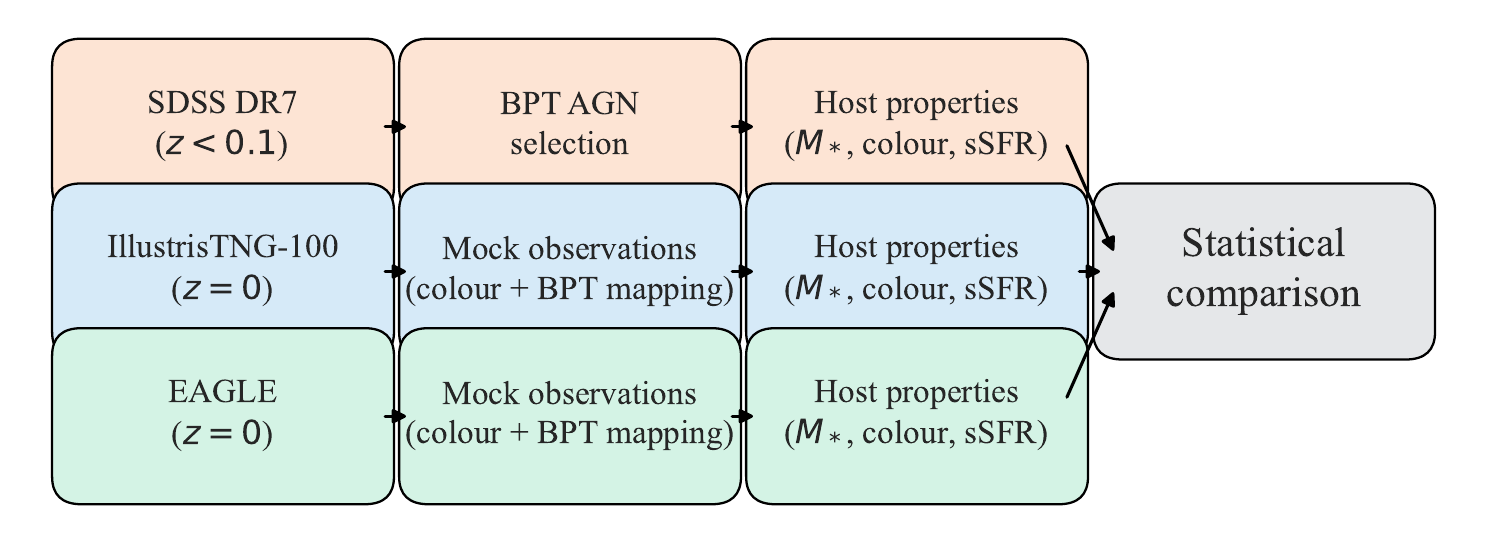}
\caption{Schematic workflow of the analysis pipeline applied to SDSS, TNG100, and EAGLE. The three parallel streams illustrate how observational SDSS DR7 galaxies ($z<0.1$) and simulated galaxies from TNG100 ($z=0$) and EAGLE ($z=0$) are processed, from catalogue ingestion through sample selection and mock-observation/visualisation steps to the extraction of host-galaxy properties. The forward-modelled BPT step is used only as an illustrative mapping for qualitative phase-space comparisons (Section~\ref{subsec:methods_bpt}); simulated galaxies are not BPT-selected as AGN. All streams converge in the right-hand block, where the statistical comparisons (Sections~\ref{sec:results} and \ref{sec:discussion}) are performed.}
\label{fig:flowchart}
\end{figure*}

All analysis steps---from catalogue ingestion through sample selection, green-valley definition, statistical calculations, and figure generation---are implemented in a single, version-controlled Python workflow based on an annotated Jupyter notebook.\footnote{The notebook is archived together with the manuscript and released via Zenodo; see the Data Availability section.} The code makes extensive use of \textsc{numpy}, \textsc{pandas}, \textsc{astropy}, \textsc{matplotlib}, and \textsc{seaborn}.\footnote{See the project \texttt{README} and \texttt{CITATION.cff} files for software citations.} All random-number draws (bootstrap resampling, control-sample matching) use explicitly recorded seeds, ensuring that the published figures and statistics can be reproduced exactly from the public data products. The overall workflow, including the parallel SDSS and simulation streams and their convergence in the final statistical comparison, is summarised schematically in Fig.~\ref{fig:flowchart}.

\section{Results}
\label{sec:results}

\subsection{Colour--mass demographics and sSFR distributions}
\label{subsec:results_colour_mass}

\begin{figure*}
\centering
\includegraphics[width=\textwidth]{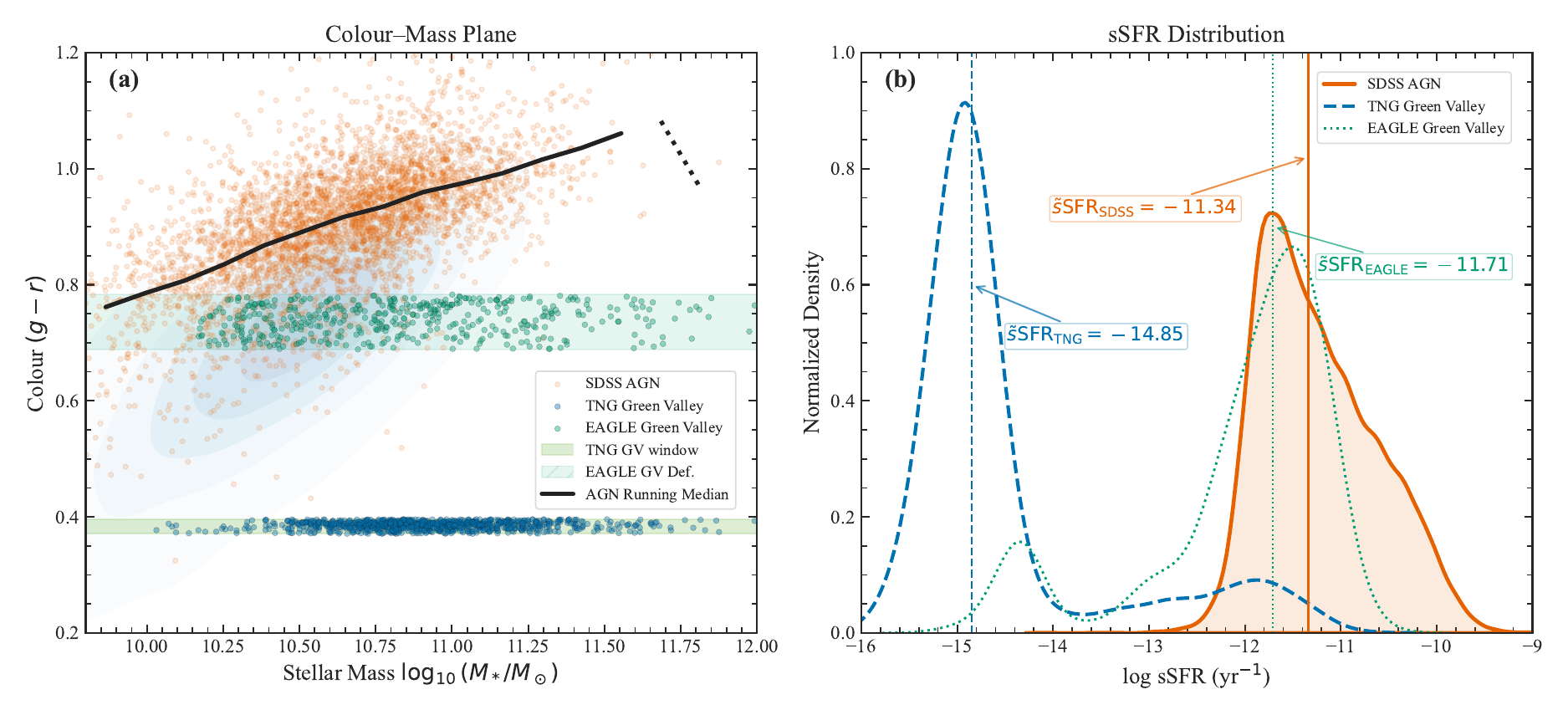}
\caption{Host-galaxy demographics of SDSS AGN compared to green-valley analogues in IllustrisTNG and EAGLE. \emph{Left:} $(g-r)$ colour versus stellar mass for SDSS DR7 galaxies (blue contours) and BPT-selected AGN hosts (orange points); for SDSS the colour is the observed-frame fibre value, whereas for the simulations it is rest-frame synthetic photometry within 30~kpc. The shaded bands indicate the 75th--95th percentile ``green-valley'' windows for TNG (light green) and EAGLE (hatched), defined separately for massive galaxies with $\log_{10}M_\star/\mathrm{M_\odot}>10$. Overplotted symbols show TNG (blue circles) and EAGLE (green diamonds) centrals lying in their respective green-valley windows. The thick black curve shows the running median colour of SDSS AGN hosts. \emph{Right:} Normalised KDEs of $\log_{10}\mathrm{sSFR}$ for SDSS AGN (orange), TNG green-valley centrals (blue dashed), and EAGLE green-valley centrals (green dotted). Vertical lines mark median values $\tilde{s}\mathrm{SFR}$ with 68~per~cent bootstrap uncertainties. TNG analogues pile up at the imposed low-sSFR pile-up corresponding to formally zero instantaneous SFR, whereas EAGLE analogues retain a broad distribution much closer to SDSS.}
\label{fig:colour_mass}
\end{figure*}

Fig.~\ref{fig:colour_mass} summarises the host-galaxy demographics of SDSS AGN hosts and their simulation analogues. In the left panel we show $(g-r)$ colour versus stellar mass. For SDSS, $(g-r)$ denotes the observed-frame fibre colour ($\mathrm{PLUGMAG}_g-\mathrm{PLUGMAG}_r$), whereas for TNG and EAGLE it denotes rest-frame synthetic photometry measured within 30~kpc; absolute colour values therefore need not match between datasets, motivating our percentile-based definition. SDSS AGN hosts occupy a broad swath of colour--mass space spanning $0.2 \lesssim (g-r) \lesssim 1.2$ and $10 \lesssim \log_{10}(M_\star/\mathrm{M_\odot}) \lesssim 12$, with a running median that tracks the upper envelope of the star-forming population but remains bluer than the classical red sequence.

Overplotted are massive centrals from TNG and EAGLE selected using their respective internal green-valley colour windows. The TNG green-valley analogues lie in a narrow $(g-r)$ interval ($\approx 0.371$--0.396) and are generally bluer than the majority of SDSS AGN hosts at fixed mass, reflecting differences in stellar populations and dust treatment rather than true physical similarity. EAGLE massive centrals exhibit a broader and intrinsically redder colour distribution; their green-valley window lies at $(g-r)\approx 0.689$--0.783 (Fig.~\ref{fig:gv_definition}), closer in absolute terms to the observed green-valley colours. However, our percentile-based definition ensures that, in all three datasets, the selected galaxies occupy the same rank-ordered portion of the massive colour distribution.

The right panel of Fig.~\ref{fig:colour_mass} shows the corresponding $\log_{10}\mathrm{sSFR}$ distributions for SDSS AGN hosts, TNG green-valley analogues, and EAGLE green-valley analogues. SDSS AGN hosts have a median $\tilde{s}\mathrm{SFR}_{\rm SDSS} \simeq -11.34$ with a broad tail toward lower sSFR. TNG green-valley analogues are dramatically offset: their distribution is sharply peaked at the imposed low-sSFR pile-up corresponding to formally zero instantaneous SFR (given the adopted $10^{-4}\,\mathrm{M_\odot\,yr^{-1}}$ SFR floor), with $\tilde{s}\mathrm{SFR}_{\rm TNG} \simeq -14.85$. EAGLE green-valley centrals occupy an intermediate regime centred at $\tilde{s}\mathrm{SFR}_{\rm EAGLE} \simeq -11.71$, only $\sim 0.3$--$0.4$~dex below SDSS, and with substantial overlap in distribution shape. Throughout what follows we use $\tilde{s}\mathrm{SFR}$ to denote these median $\log_{10}\mathrm{sSFR}$ values for each sample.

Two-sample K--S tests confirm these impressions: for the sSFR distributions shown in Fig.~\ref{fig:colour_mass}, SDSS AGN hosts and TNG analogues differ at the $D\approx 0.90$ level, whereas the SDSS--EAGLE comparison yields $D\approx 0.315$. The TNG deficit in sSFR is therefore both large and statistically robust.

\subsection{sSFR--mass plane}
\label{subsec:results_ssfr_mass}

\begin{figure*}
\centering
\includegraphics[width=\textwidth]{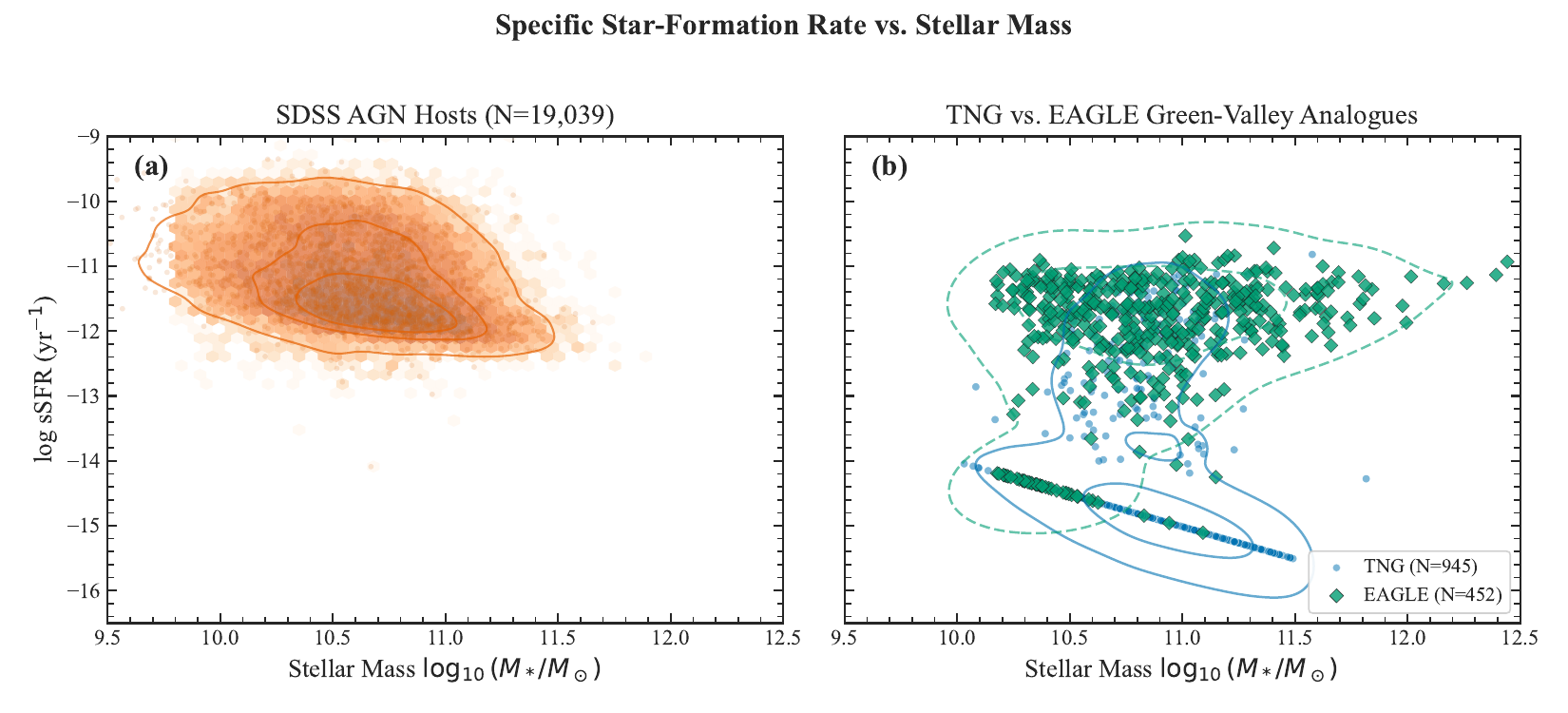}
\caption{Specific star-formation rate versus stellar mass. \emph{Left:} SDSS BPT-selected AGN hosts at $z<0.1$ shown as orange points (down-sampled for clarity) over a hexagonal density map with overlaid KDE contours. \emph{Right:} TNG100 (blue circles) and EAGLE (green diamonds) green-valley centrals selected via simulation-specific colour percentiles. Contours show the core of each distribution. SDSS hosts occupy a broad, partially quenched locus. TNG analogues collapse onto a low-sSFR pile-up at $\log_{10}\mathrm{sSFR}\approx -15.5$, whereas EAGLE analogues span a wide range of intermediate sSFRs that more closely resembles SDSS.}
\label{fig:ssfr_mass}
\end{figure*}

Fig.~\ref{fig:ssfr_mass} examines the structure of the sSFR--mass plane. SDSS AGN hosts populate a broad, partially quenched locus that lies $\sim 1$~dex below the star-forming main sequence but well above fully passive levels. Their distribution shows a gradual decline in sSFR with increasing mass, consistent with a mix of transitional and already-quenched hosts.

By contrast, TNG green-valley analogues cluster in a narrow band at $\log_{10}\mathrm{sSFR}\approx -15.5$, independent of stellar mass, with very few objects inhabiting the intermediate $-13\lesssim \log_{10}\mathrm{sSFR}\lesssim -11$ range. This nearly horizontal low-sSFR pile-up reflects the kinetic-mode AGN feedback in TNG, which rapidly suppresses cold gas accretion and star formation once activated.

EAGLE green-valley centrals, on the other hand, span a broad range $-13 \lesssim \log_{10}\mathrm{sSFR}\lesssim -10$ with a smoothly declining upper envelope that qualitatively parallels the SDSS AGN locus. This behaviour indicates a more gradual quenching pathway in which galaxies pass through a continuum of intermediate sSFR states rather than collapsing directly onto a floor.

\subsection{Cumulative distributions and median properties}
\label{subsec:results_cdfs}

\begin{figure*}
\centering
\includegraphics[width=\textwidth]{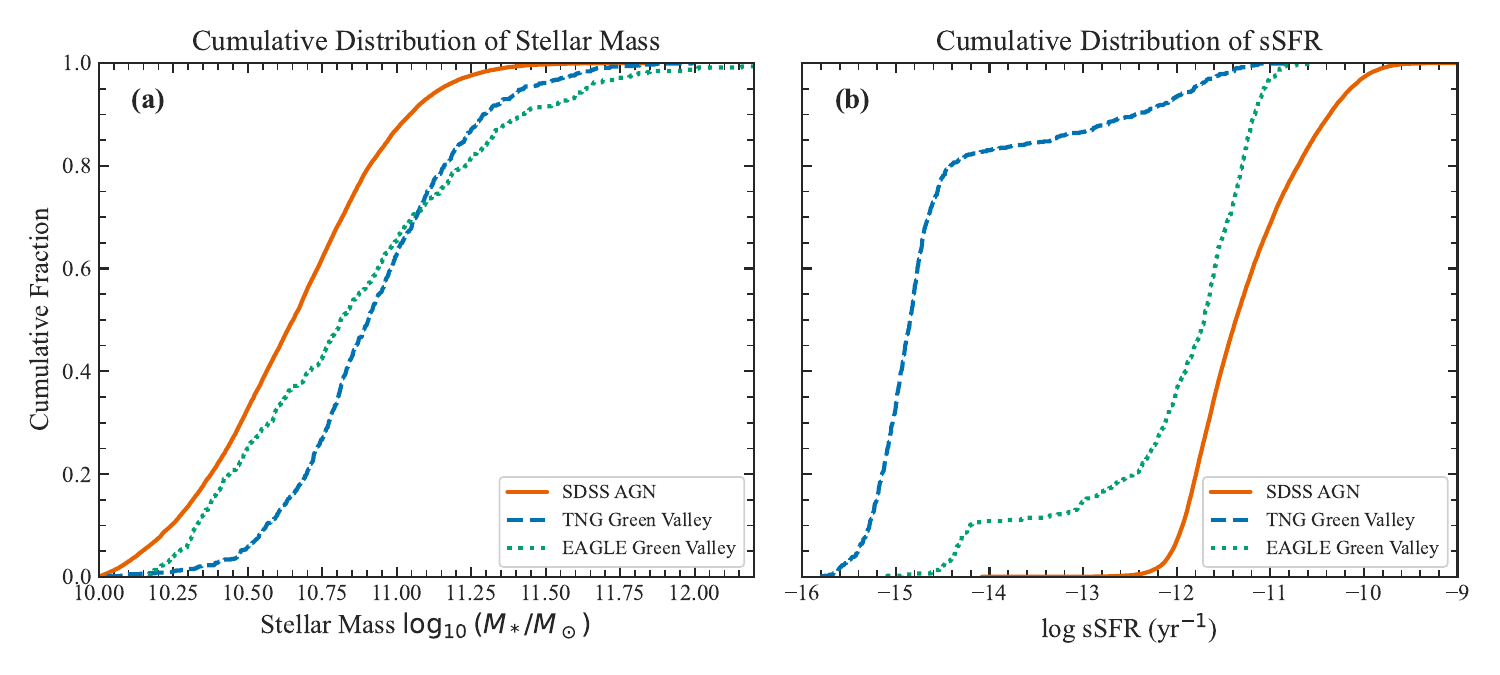}
\caption{Cumulative distributions of stellar mass (left) and sSFR (right) for massive ($\log_{10}M_\star>10$) SDSS AGN hosts (orange), TNG green-valley centrals (blue dashed), and EAGLE green-valley centrals (green dotted). All quantities are measured within 30~kpc apertures in the simulations and derived from aperture-corrected totals in SDSS. TNG green-valley analogues are both more massive and far more quenched than SDSS AGN hosts, while EAGLE analogues are closer in both mass and sSFR distribution.}
\label{fig:cdfs}
\end{figure*}

Fig.~\ref{fig:cdfs} shows cumulative distributions of stellar mass and sSFR for the mass-limited samples ($\log_{10}M_\star > 10$). The stellar-mass CDFs reveal that both simulations produce green-valley samples that are more massive, on average, than the SDSS AGN hosts. Bootstrapped medians for $\log_{10}M_\star$ are
\begin{align}
  \tilde{M}_\star^{\rm SDSS} &= 10.652^{+0.004}_{-0.003}, \\
  \tilde{M}_\star^{\rm TNG}  &= 10.905^{+0.010}_{-0.013}, \\
  \tilde{M}_\star^{\rm EAGLE} &= 10.812^{+0.031}_{-0.015},
\end{align}
in units of $\log_{10}(\mathrm{M_\odot})$ (see Appendix~\ref{sec:appendix_bootstrap}). K--S tests confirm that all three mass distributions are statistically distinct, but the EAGLE CDF is qualitatively closer in shape to that of SDSS AGN hosts than the TNG CDF.

The sSFR CDFs show even more pronounced differences: TNG rises extremely steeply at the low-sSFR pile-up, indicating that nearly all green-valley analogues are effectively passive. EAGLE exhibits a more gradually rising CDF that more closely tracks the SDSS curve, offset by only a few tenths of a dex in median sSFR. The bootstrapped median $\tilde{s}\mathrm{SFR}$ values for the mass-limited samples are
\begin{align}
  \tilde{s}\mathrm{SFR}^{\rm SDSS} &= -11.369^{+0.007}_{-0.005}, \\
  \tilde{s}\mathrm{SFR}^{\rm TNG}  &= -14.850^{+0.015}_{-0.014}, \\
  \tilde{s}\mathrm{SFR}^{\rm EAGLE} &= -11.707^{+0.029}_{-0.039}.
\end{align}
Thus, the TNG green-valley population is offset from SDSS by $\sim 3.5$~dex in sSFR, whereas EAGLE is offset by only $\sim 0.3$--$0.4$~dex.

\subsection{Green-valley occupancy fractions}
\label{subsec:results_gvfrac}

\begin{figure}
\centering
\includegraphics[width=\columnwidth]{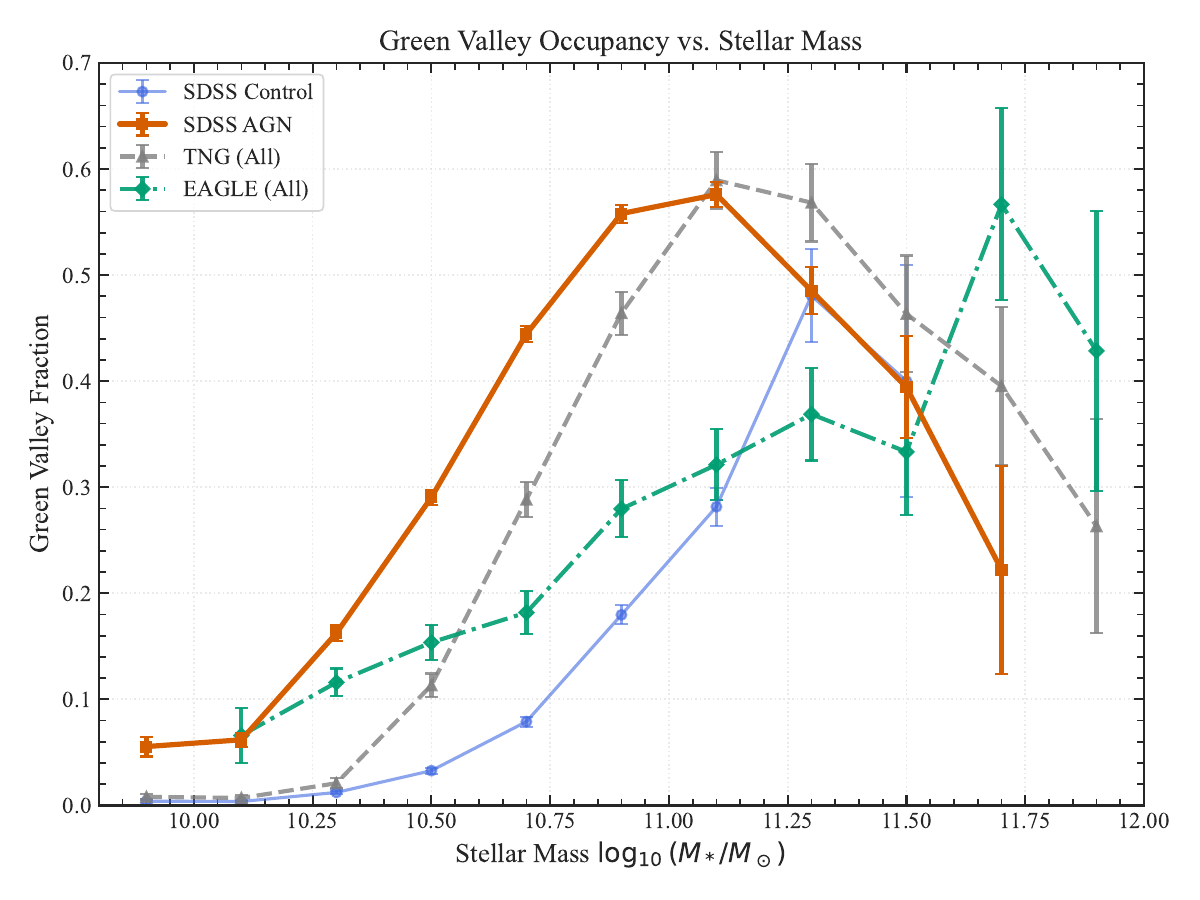}
\caption{Green-valley occupancy fraction as a function of stellar mass. The fraction of galaxies in the percentile-defined green valley is shown for SDSS AGN hosts (orange squares), a matched SDSS star-forming control sample (blue circles), all TNG100 centrals (grey triangles), and all EAGLE centrals (green diamonds). Error bars are binomial $1\sigma$ uncertainties; bins with fewer than 10 galaxies are omitted. AGN hosts show an enhanced green-valley fraction relative to the control sample, peaking near $\log_{10}M_\star\simeq 11$, while TNG produces an even higher occupancy fraction dominated by already-quenched systems. EAGLE yields a lower and more gradually varying occupancy.}
\label{fig:gv_occupancy}
\end{figure}

Fig.~\ref{fig:gv_occupancy} presents the fraction of galaxies residing in the green valley as a function of stellar mass. For SDSS we show both the AGN hosts and the matched star-forming control sample; for the simulations we show all centrals in TNG and EAGLE.

The control sample exhibits a modest green-valley occupancy that rises from a few per~cent at $10^{10}\,\mathrm{M_\odot}$ to $\sim 30$~per~cent near $10^{11}\,\mathrm{M_\odot}$ before declining. AGN hosts show a significantly higher occupancy at fixed mass, peaking at $\sim 50$--60~per~cent near $\log_{10}M_\star\approx 11$, consistent with AGN preferentially inhabiting transitional systems.

TNG predicts a sharply rising green-valley fraction with mass that exceeds the SDSS AGN occupancy at high masses, approaching $\sim 60$~per~cent. This high fraction is driven largely by already-quenched galaxies that happen to fall into the colour percentile window, as demonstrated by their extremely low sSFRs. EAGLE yields a more modest and broader occupancy profile, peaking at $\sim 30$--40~per~cent and remaining below the SDSS AGN hosts over most of the mass range. These trends reflect the contrasting quenching efficiencies and timescales of the two feedback models.

\subsection{Robustness to green-valley definition}
\label{subsec:results_robust}

\begin{figure}
\centering
\includegraphics[width=\columnwidth]{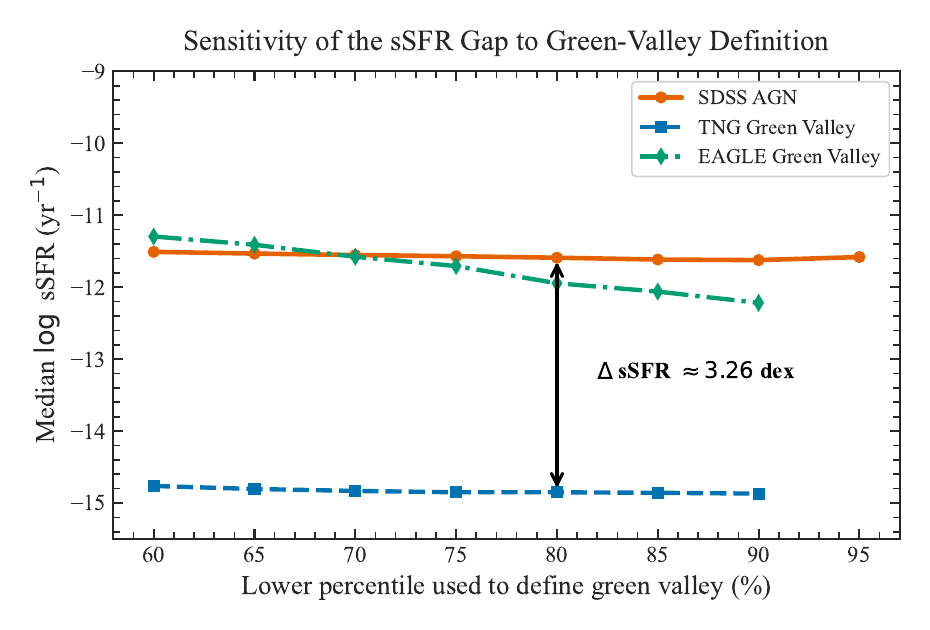}
\caption{Sensitivity of the median sSFR to the adopted green-valley colour percentile. The upper percentile is fixed at 95~per~cent, while the lower percentile is varied from 60 to 90~per~cent. Points show the median $\tilde{s}\mathrm{SFR}$ of galaxies in the resulting green-valley windows for SDSS AGN hosts (orange), TNG100 centrals (blue), and EAGLE centrals (green). TNG medians remain fixed at the low-sSFR pile-up, whereas SDSS and EAGLE medians change only mildly, with EAGLE consistently tracking SDSS much more closely.}
\label{fig:percentile_sweep}
\end{figure}

To assess whether our conclusions depend sensitively on the chosen colour percentiles, we recompute the median sSFR of the green-valley samples as a function of the lower percentile used to define the colour window, keeping the upper percentile fixed at 95~per~cent (Fig.~\ref{fig:percentile_sweep}). In all three datasets we consider lower bounds ranging from the 60th to the 90th percentiles.

Across this entire range, TNG green-valley analogues remain pinned to the low-sSFR pile-up at $\tilde{s}\mathrm{SFR}_{\rm TNG}\approx -14.8$, indicating that the extreme sSFR deficit is not an artefact of the precise green-valley definition. The SDSS AGN median varies only weakly with percentile, staying between $-11.3$ and $-11.6$, while EAGLE tracks SDSS within $\sim 0.3$--$0.5$~dex for all choices. The large SDSS--TNG discrepancy and the much closer SDSS--EAGLE alignment are therefore robust to reasonable variations in the green-valley selection.

\subsection{Bootstrap K--S distributions}
\label{subsec:results_bootstrap_ks}

\begin{figure}
\centering
\includegraphics[width=\columnwidth]{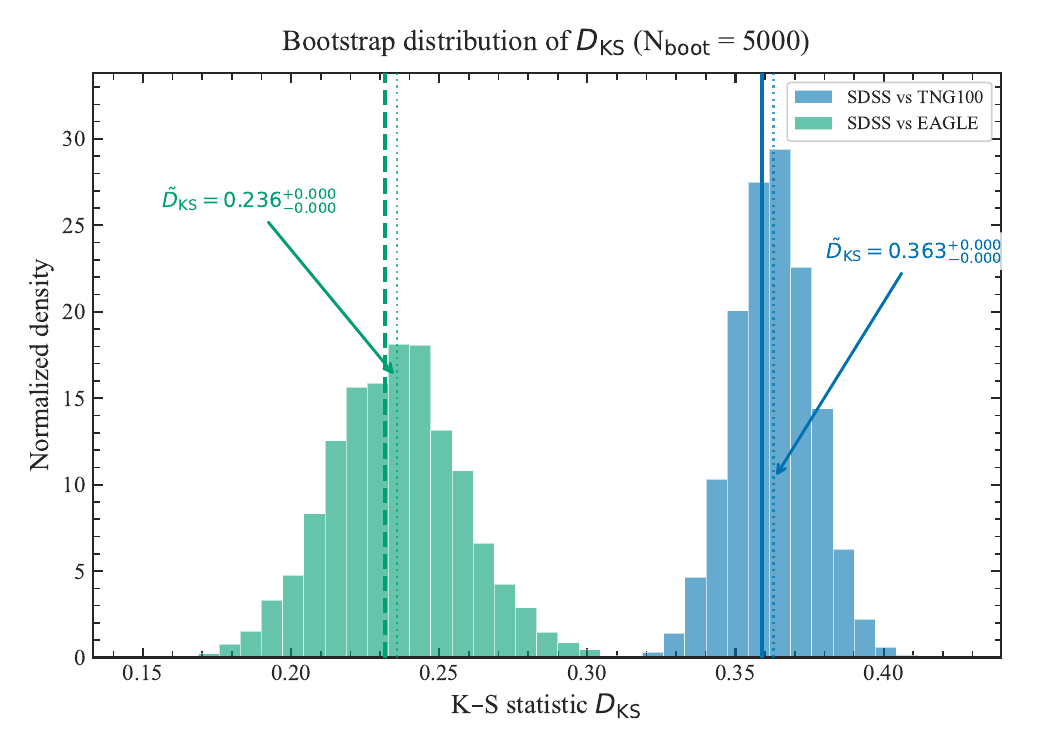}
\caption{Bootstrap distributions of the K--S statistic $D_{\rm KS}$ comparing the stellar-mass distributions of SDSS AGN hosts with TNG100 (blue) and EAGLE (green) green-valley centrals. Solid and dashed vertical lines mark the observed $D_{\rm KS}$ values for the original samples; faint dotted vertical lines indicate the bootstrap medians $\tilde{D}_{\rm KS}$ for each distribution, with arrow annotations reporting the medians together with the central 68~per~cent ranges of the bootstrap distributions. In practice these ranges are extremely narrow at the precision shown, so the quoted intervals may appear as $0.000$ when rounded to three decimal places. TNG systematically yields larger $D_{\rm KS}$ values, indicating a more pronounced mismatch in stellar-mass distribution relative to SDSS than EAGLE.}
\label{fig:bootstrap_ks}
\end{figure}

Fig.~\ref{fig:bootstrap_ks} shows the bootstrap distributions of the K--S statistic $D_{\rm KS}$ comparing the stellar-mass distributions of SDSS AGN hosts with the TNG and EAGLE green-valley samples. For each of $N_{\rm boot}=5000$ iterations we resample the SDSS AGN host masses and the simulation masses with replacement and compute the two-sample K--S statistic for SDSS--TNG and SDSS--EAGLE.

The resulting distributions are narrow and well-separated. The K--S statistics measured from the original samples are $D_{\rm KS}=0.232$ for SDSS--EAGLE and $D_{\rm KS}=0.359$ for SDSS--TNG, and the bootstrap distributions cluster tightly around these values. The observed K--S statistics for the original samples lie well within their respective bootstrap distributions. This experiment demonstrates that, even after accounting for sampling variance and the differing sample sizes, the TNG green-valley masses are substantially more discrepant from the SDSS AGN host masses than the EAGLE masses are.

\subsection{Forward-modelled BPT classifications}
\label{subsec:results_bpt}

\begin{figure}
\centering
\includegraphics[width=\columnwidth]{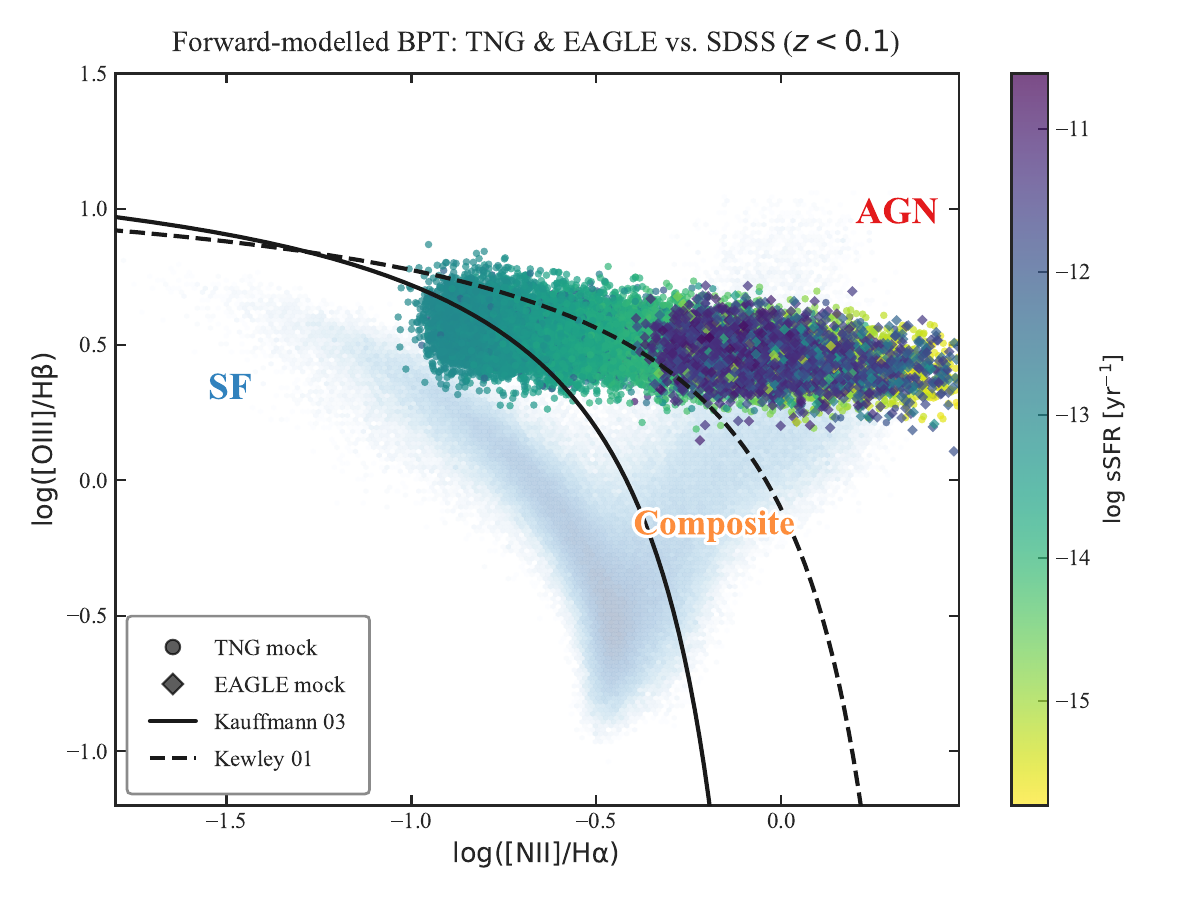}
\caption{Forward-modelled BPT diagram for TNG100 and EAGLE galaxies compared to SDSS DR7. The blue hexbin background shows the SDSS distribution at $z<0.1$. Points indicate mock TNG (circles) and EAGLE (diamonds) galaxies with emission-line ratios derived from a simple mass--metallicity mapping. Colours encode $\log_{10}\mathrm{sSFR}$, with lower sSFR in darker shades. Solid and dashed curves show the \citet{Kauffmann2003} and \citet{Kewley2001} demarcation lines. TNG analogues occupy a narrow locus of low-sSFR systems, while EAGLE analogues span a broader range of sSFR and more closely track the SDSS star-forming and composite loci.}
\label{fig:bpt_mock}
\end{figure}

Fig.~\ref{fig:bpt_mock} presents the forward-modelled BPT diagram for TNG and EAGLE galaxies, overlaid on the SDSS DR7 hexbin background. The mock points broadly trace the observed star-forming sequence and its turn-off toward higher [N\,\textsc{ii}]/H$\alpha$ at high metallicity. When colour-coded by sSFR, the mock TNG points are predominantly low-sSFR systems, consistent with the severe quenching seen in the sSFR distributions. EAGLE points span a wider range of sSFR, including many galaxies with relatively high sSFR that lie near the star-forming and composite regions of the diagram.

Using the \citet{Kewley2001} curve as a formal boundary in the forward model, we find that a substantial fraction of mock points scatter above the line under this mapping. These fractions should not be interpreted quantitatively, given the simplicity of the forward model and the lack of explicit AGN emission; instead, they serve only to illustrate that TNG and EAGLE populate different parts of BPT phase space once their host properties are taken into account.

\section{Discussion}
\label{sec:discussion}

Before interpreting these trends physically, we reiterate that our SDSS comparison sample is emission-line-selected (BPT-classified) whereas the simulation samples are colour-selected centrals; accordingly, the comparison is best read as a test of quenching pathways in the transitional colour regime rather than a one-to-one reproduction of the SDSS AGN selection function. Although SDSS hosts are selected by nebular activity and simulated hosts by colour, this comparison asks whether the colour-defined ``transitional'' phase in simulations corresponds to the host-galaxy regime occupied by emission-line AGN in the local Universe. The analyses above paint a coherent picture of two qualitatively different quenching pathways in contemporary cosmological simulations.

In IllustrisTNG, green-valley analogues selected via internal colour percentiles are already fully quenched in sSFR. Once the kinetic AGN feedback mode activates, galaxies are driven rapidly to extremely low sSFRs with little time spent in intermediate states. This leads to an artificial ``pile-up'' of passive systems in the colour-defined green valley and a strong mismatch with the sSFR distribution of SDSS AGN hosts. A particularly stark manifestation of this behaviour is that, when we apply the TNG green-valley colour window to the matched SDSS pure-AGN host catalogue, only 6 out of 19\,039 hosts ($\simeq 0.03$~per~cent) have fibre colours within the same $g-r$ interval.\footnote{These numbers are obtained directly from the matched SDSS AGN catalogue by counting hosts with $g-r$ colours in the TNG 75th--95th percentile window; see the accompanying analysis notebook for details.} In other words, the overlap in colour space between the observed AGN hosts and the TNG green-valley band is effectively negligible. Physically, this behaviour is consistent with the low-Eddington kinetic wind mode depositing momentum and energy into the circumgalactic medium so efficiently that it clears or overheats halo gas, cutting off the inflow of fresh fuel and driving galaxies almost directly onto the low-sSFR pile-up. The TNG kinetic mode clearly succeeds in producing quenched high-mass galaxies and a pronounced colour bimodality, but our results suggest that it does so in a way that is too efficient and too abrupt: the transition is so rapid that the simulated population virtually evacuates the green valley of star-forming gas before galaxies can redden through an extended transitional phase comparable to that traced by SDSS AGN hosts.

In EAGLE, the thermal AGN feedback scheme acts more gently on the star-forming gas, heating and stabilising it without instantaneously removing or shutting down the supply. Accretion energy is accumulated and then released in stochastic thermal events, which tend to puff up and heat the gas rather than immediately ejecting it from the halo. As a result, EAGLE green-valley centrals span a broad continuum of sSFRs, with a median $\tilde{s}\mathrm{SFR}$ only a few tenths of a dex below SDSS AGN hosts, and many galaxies lingering at intermediate sSFRs. This behaviour is naturally interpreted as a ``slow-fade'' or strangulation-like quenching pathway in which the fuel supply is gradually starved rather than catastrophically destroyed. The green-valley occupancy fractions and mass distributions also resemble SDSS more closely, although both simulations still tend to produce somewhat more massive transitional centrals than seen in the observed AGN host sample.

Systematic uncertainties in SFR and colour calibration---including fibre versus aperture effects, dust modelling, and stellar-population synthesis---are unlikely to explain the extreme $\sim 3.5$~dex TNG--SDSS discrepancy in $\tilde{s}\mathrm{SFR}$. Such systematics are typically at the few-tenths-of-a-dex level in sSFR and colour for the datasets considered here. We also note (Section~\ref{subsec:data_sdss}) that the SDSS BPT selection imposes an emission-line detectability requirement ($S/N>3$), which filters out very weak-lined systems and therefore affects completeness at very low sSFR; however, the observed multi-dex offset between SDSS and TNG is orders of magnitude larger than can plausibly be explained by this bias alone. The much smaller $\sim 0.3$--$0.4$~dex offset between EAGLE and SDSS is comfortably within the expected scale of these systematics, lending further weight to the interpretation that the main differences we observe are driven by the underlying feedback physics, not by observational calibration alone.

Our results complement and extend previous work highlighting quenching tensions in large-volume simulations. Several studies have found that TNG tends to over-quench massive centrals and under-predict the prevalence of star-forming galaxies in massive haloes at $z\sim 0$ \citep[e.g.][]{Donnari2019}, while EAGLE is somewhat more successful at maintaining a population of mildly star-forming massive galaxies \citep[e.g.][]{Schaye2015,Trayford2016}. Those earlier comparisons were largely phrased in terms of the global quenched fraction and related summary statistics: they showed that TNG can reproduce the overall fraction of passive galaxies at fixed stellar mass---the correct \emph{destination} in colour--mass space. By focusing specifically on the green valley and on BPT-selected AGN hosts, our analysis instead isolates the \emph{quenching pathway}. The $\tilde{s}\mathrm{SFR}$ distributions demonstrate that, while TNG reaches the right destination, the \emph{journey} is too abrupt: the kinetic mode drives galaxies almost instantaneously from the star-forming locus to extreme quiescence, leaving very few objects in the intermediate regime that dominates the SDSS AGN host population. EAGLE, by contrast, reproduces not only a plausible quenched fraction but also a broad, slowly declining $\tilde{s}\mathrm{SFR}$ distribution through the green valley that more closely matches the observed journey of low-redshift AGN hosts.

\section{Conclusions}
\label{sec:conclusions}

We have carried out a detailed, reproducible comparison of low-redshift AGN host galaxies in SDSS DR7 with green-valley analogue galaxies selected from the TNG100-1 and EAGLE Ref-L0100N1504 simulations. Because we do not impose an explicit AGN selection in the simulations, the TNG and EAGLE samples should be viewed as colour-selected green-valley analogues; the comparison therefore constrains how different feedback models populate (or evacuate) intermediate-sSFR states relative to the host-galaxy properties of BPT-selected AGN. By defining the green valley via internal colour percentiles in each dataset and matching stellar-mass thresholds, we minimise calibration differences and focus on relative quenching behaviour.

Our main conclusions are:
\begin{enumerate}
    \item \textbf{TNG100 over-quenches green-valley analogues.} TNG green-valley centrals with $\log_{10}M_\star>10$ have median sSFRs around $\log_{10}\mathrm{sSFR}\simeq -14.85$, i.e.\ $\sim 3.5$~dex below SDSS AGN hosts (median $\log_{10}\mathrm{sSFR}\simeq -11.34$ for the matched sample). Their sSFR distribution is dominated by a sharp pile-up at the imposed analysis floor corresponding to formally zero instantaneous SFR, indicating that the kinetic AGN feedback mode acts as an almost binary switch that leaves virtually no galaxies in the intermediate sSFR regime characteristic of observed green-valley AGN hosts.
    \item \textbf{EAGLE reproduces a realistic transitional regime.} EAGLE green-valley centrals exhibit a broad sSFR distribution centred at $\log_{10}\mathrm{sSFR}\simeq -11.71$, only $\sim 0.3$--$0.4$~dex below SDSS. Their sSFR and mass distributions, as well as the green-valley occupancy fractions, qualitatively match the SDSS AGN hosts far better than TNG's do. The EAGLE feedback model thus supports a more gradual quenching pathway consistent with a ``slow-fade'' of star formation.
    \item \textbf{AGN prefer the green valley, but simulations differ in occupancy.} In SDSS, AGN hosts show an enhanced green-valley occupancy fraction relative to mass- and magnitude-matched star-forming controls, peaking at $\sim 50$--$60$~per~cent near $M_\star \sim 10^{11}\,\mathrm{M_\odot}$. TNG predicts an even higher occupancy fraction, driven largely by already-quenched systems that remain in the green-valley colour window, while EAGLE yields a lower, broadly distributed occupancy that is more in line with the SDSS trends than TNG.
    \item \textbf{The qualitative conclusions are robust to selection choices.} Varying the green-valley percentile definition over a wide range and performing bootstrap K--S experiments do not alter the qualitative picture: TNG remains strongly over-quenched, while EAGLE stays close to SDSS in sSFR behaviour.
\end{enumerate}

These findings imply that AGN feedback models tuned to reproduce global statistics such as the quenched fraction and stellar mass function may still differ substantially in how galaxies traverse the green valley. Our comparison suggests that quenching in the local Universe is more consistent with a gradual suppression of star formation, as realised in EAGLE, than with the nearly instantaneous shutdown implied by TNG's kinetic feedback channel.

\subsection*{Future work}

The framework developed here can be extended in several directions. Applying the same percentile-based methodology to other large simulations (e.g.\ SIMBA, Horizon-AGN, IllustrisTNG300, \citealt{Dave2019,Dubois2016,Pillepich2018}) would help determine whether gradual or binary quenching behaviour is generic to particular classes of feedback prescriptions. Incorporating more detailed forward-modelling of emission-line ratios---for example, via photoionisation modelling coupled to resolved gas properties---would refine the comparison to BPT-selected AGN, especially in the composite regime. Finally, tracing green-valley analogues back through time in simulations could link the present-day sSFR distributions to the long-term accretion histories of their central black holes, providing a more direct test of AGN-driven quenching scenarios.

\section*{Data Availability}
\label{sec:data_availability}

The SDSS DR7 galaxy catalogues and MPA--JHU emission-line measurements are available from the MPA Garching SDSS DR7 portal.\footnote{\url{https://wwwmpa.mpa-garching.mpg.de/SDSS/DR7/Data/}} The IllustrisTNG100-1 simulation data are publicly accessible via the TNG project database.\footnote{\url{https://www.tng-project.org/data/downloads/TNG100-1/}} EAGLE Ref-L0100N1504 simulation outputs can be accessed through the EAGLE SQL interface.\footnote{\url{http://virgodb.dur.ac.uk:8080/Eagle/}} All data processing scripts,
statistical analysis code, and figure-generation notebooks used in this study are openly available in the project
GitHub repository.\footnote{\url{https://github.com/TshapedAsh/Green-Valley-AGN-SDSS-TNG/}}
The complete processed datasets and all final figure files are archived on Zenodo under DOI \href{https://doi.org/10.5281/zenodo.17917468}{10.5281/zenodo.17917468}.

\section*{Acknowledgements}

I thank the SDSS, IllustrisTNG, and EAGLE collaboration teams for making their data publicly available, and acknowledge the developers of the open-source Python scientific ecosystem, including \textsc{numpy}, \textsc{scipy}, \textsc{pandas}, \textsc{astropy}, \textsc{matplotlib}, and \textsc{seaborn}. Parts of this work used AI-assisted coding and reproducibility tools; all scientific choices, validation, and interpretation were performed by the author.

\bibliographystyle{mnras}
\bibliography{references}

\appendix

\section{Emission-line classification of SDSS galaxies}
\label{sec:appendix_bpt}

\begin{figure}
\centering
\includegraphics[width=\columnwidth]{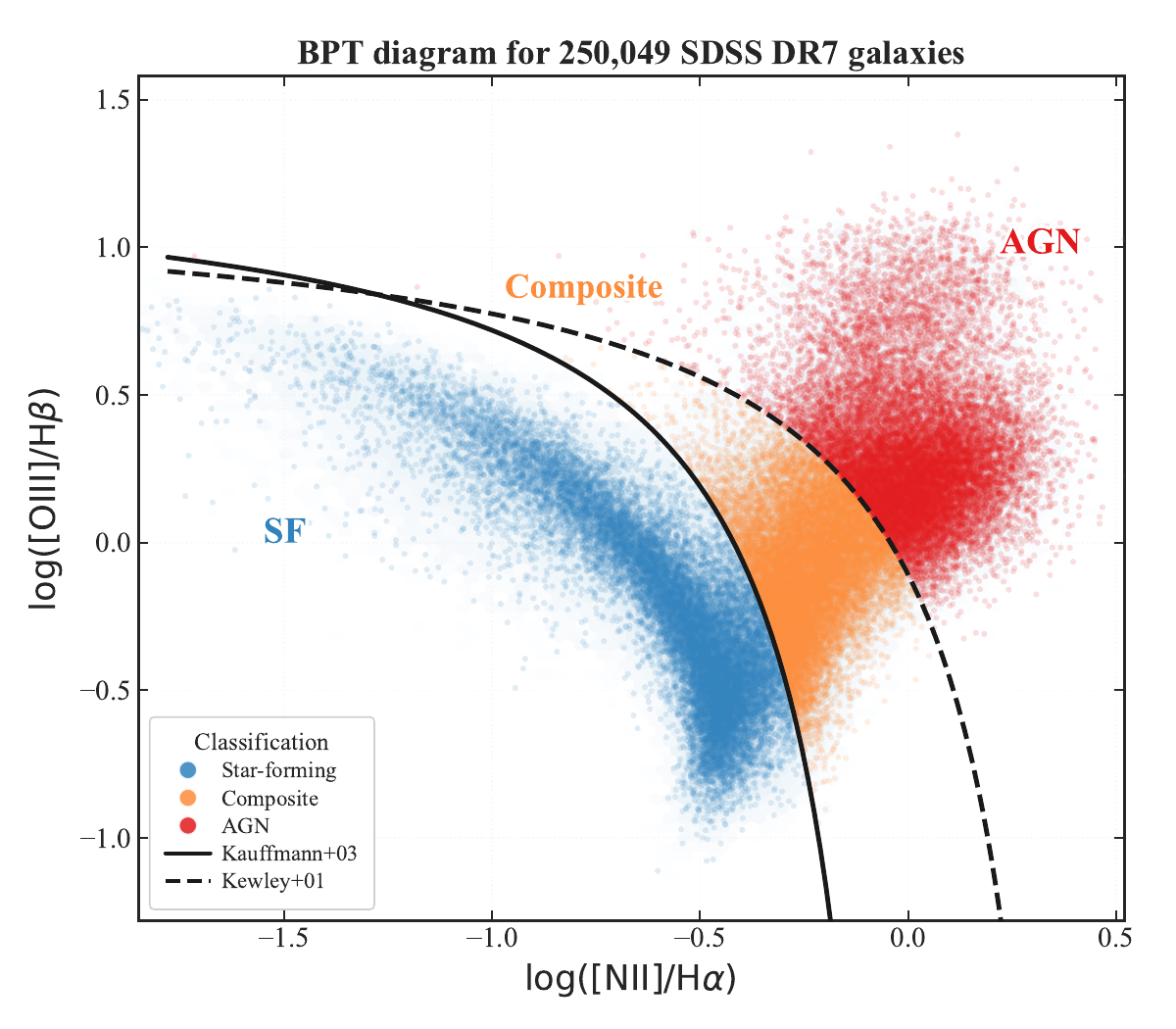}
\caption{Classical BPT diagram for SDSS DR7 galaxies at $z<0.1$. The blue background shows the full emission-line sample; coloured points indicate randomly down-sampled star-forming (blue), composite (orange), and AGN (red) galaxies. Solid and dashed curves mark the \citet{Kauffmann2003} and \citet{Kewley2001} demarcations. Only galaxies above the Kewley line enter our pure-AGN sample.}
\label{fig:bpt_sdss}
\end{figure}

Fig.~\ref{fig:bpt_sdss} presents the classical BPT diagram for the parent SDSS DR7 emission-line sample after quality cuts ($N\simeq 2.5\times 10^5$ galaxies). The blue hexbin background shows the full population density. Overplotted are randomly down-sampled star-forming, composite, and AGN subsamples colour-coded by their BPT class. The \citet{Kauffmann2003} and \citet{Kewley2001} curves demarcate the star-forming, composite, and AGN/LINER regions; only galaxies above the Kewley line enter our pure-AGN host sample in the main analysis.

\section{Green-valley colour definition in TNG and EAGLE}
\label{sec:appendix_gv}

\begin{figure}
\centering
\includegraphics[width=\columnwidth]{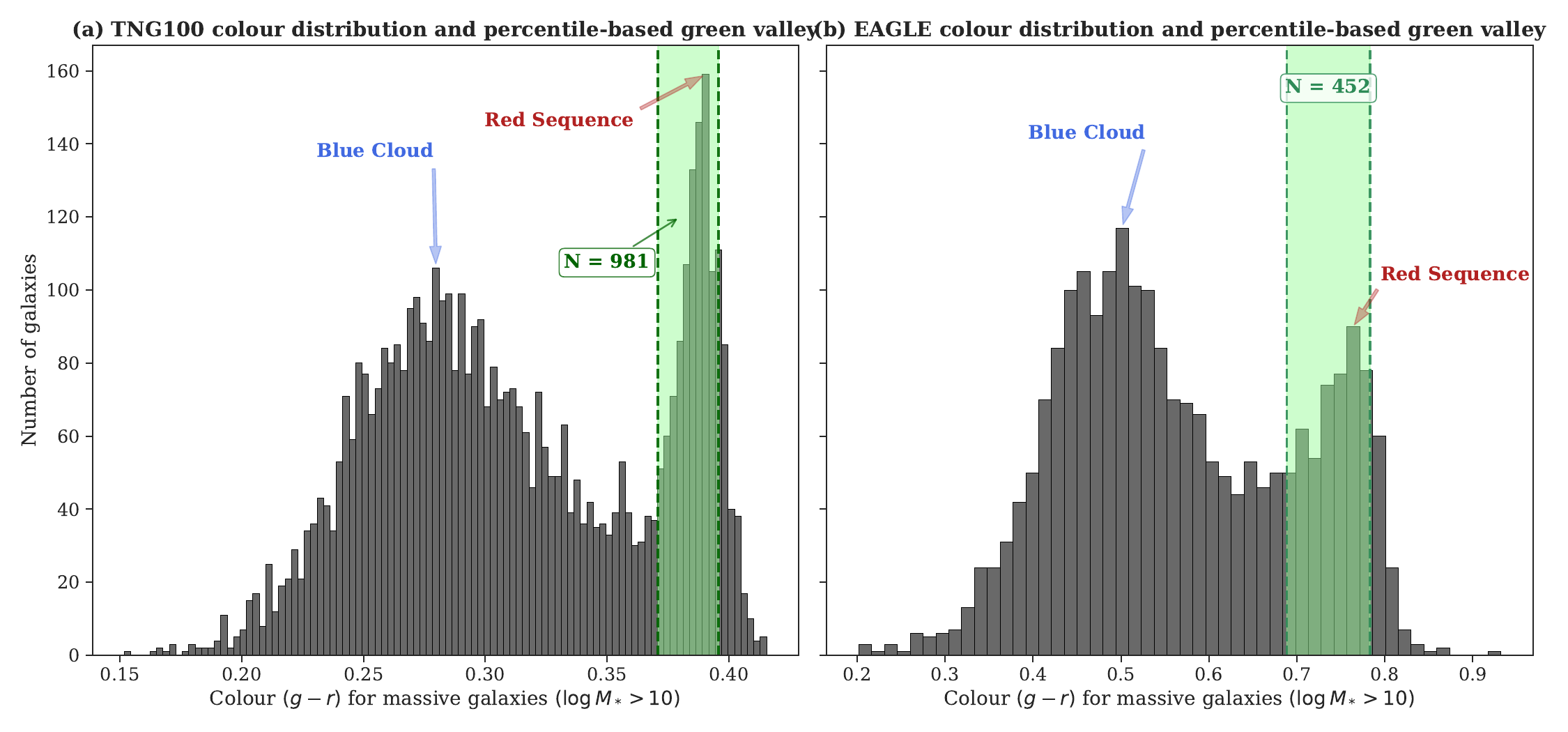}
\caption{Rest-frame $(g-r)$ colour distributions for massive centrals ($\log_{10}M_\star>10$) in TNG100 (left) and EAGLE (right). Grey histograms show the full massive-galaxy populations; shaded bands indicate the 75th--95th percentile ``green-valley'' windows used to define simulation-specific analogues. Annotated counts give the number of centrals in each window. Although TNG and EAGLE differ strongly in their absolute colour distributions, the percentile-based selection always isolates the reddest $\sim 20$~per~cent of massive galaxies short of the red sequence.}
\label{fig:gv_definition}
\end{figure}

Fig.~\ref{fig:gv_definition} shows histograms of rest-frame $(g-r)$ colour for massive centrals ($\log_{10}M_\star>10$) in TNG100 and EAGLE. In each panel the shaded green band marks the 75th--95th percentile range used to define the simulation-specific green valley, and the annotated $N$ values give the number of centrals in this interval. The TNG distribution is relatively narrow and blue, whereas the EAGLE distribution is broader and redder; a K--S test yields $D\approx 0.90$ for the TNG--EAGLE colour distributions. The percentile-based definition ensures that we always select the same \emph{relative} part of each distribution despite these absolute differences.

\section{Stellar-mass distributions}
\label{sec:appendix_mass}

\begin{figure}
\centering
\includegraphics[width=\columnwidth]{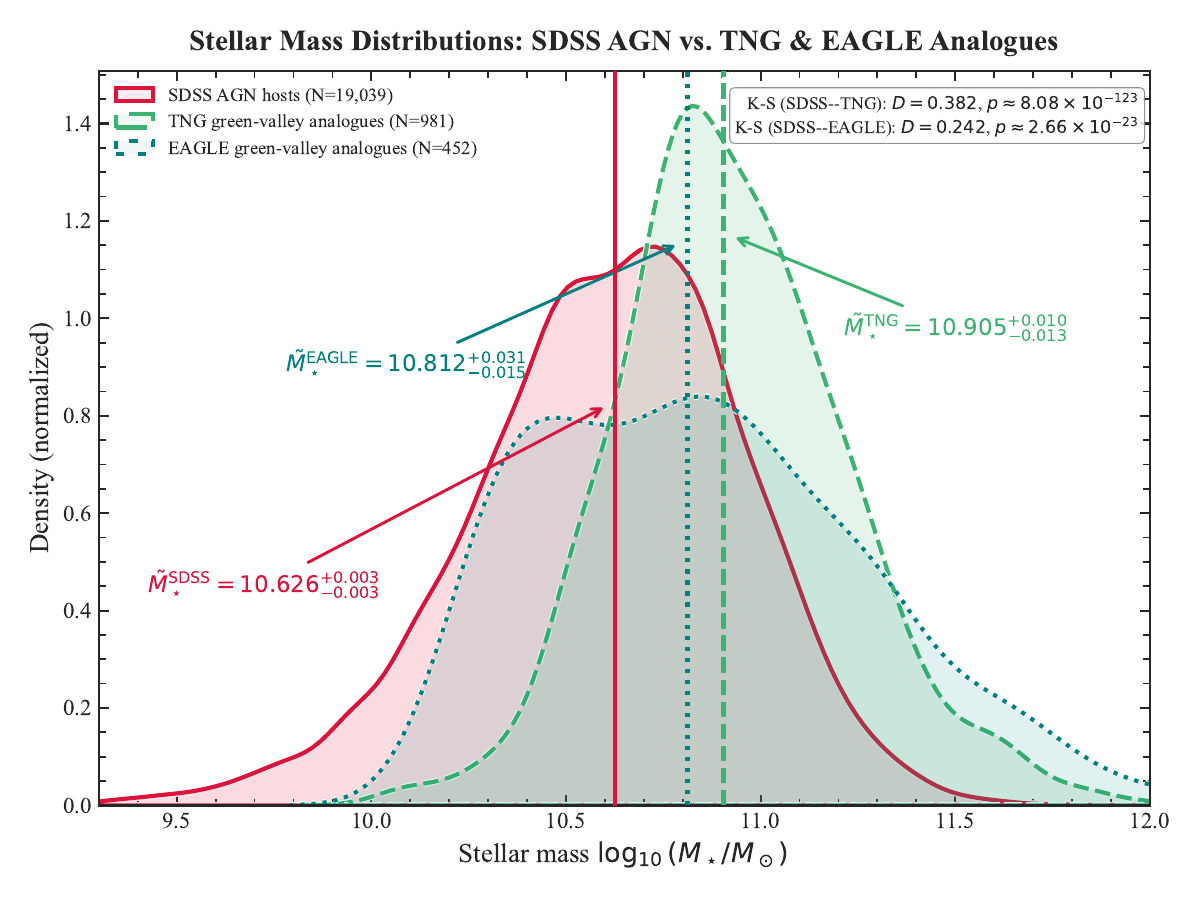}
\caption{Stellar-mass distributions of SDSS AGN hosts (solid orange), TNG100 green-valley centrals (dashed blue), and EAGLE green-valley centrals (dotted green). Vertical lines and annotations mark median masses with 68~per~cent bootstrap confidence intervals. Both simulations favour somewhat higher masses than SDSS, but EAGLE is qualitatively closer in distribution shape.}
\label{fig:mass_dist}
\end{figure}

Fig.~\ref{fig:mass_dist} compares the stellar-mass distributions of SDSS AGN hosts, TNG100 green-valley analogues, and EAGLE green-valley analogues. KDEs are shown for each sample, with vertical lines marking the bootstrapped median and annotated 68~per~cent confidence intervals. K--S tests indicate that both simulations differ significantly from SDSS; in Fig.~\ref{fig:mass_dist} the SDSS--EAGLE mass distribution is closer in shape than SDSS--TNG (as quantified by the corresponding K--S $D$ values).

\section{Bootstrapped medians for mass and sSFR}
\label{sec:appendix_bootstrap}

This appendix summarises bootstrapped medians and 68~per~cent confidence intervals for the stellar-mass and sSFR distributions of massive SDSS AGN hosts and simulation analogues. For $\log_{10}M_\star>10$ we find the values listed in Table~\ref{tab:bootstrap_stats}. These numbers underpin the qualitative conclusions drawn in Sections~\ref{sec:results} and \ref{sec:conclusions}.

\begin{table}
    \centering
    \caption{Bootstrapped medians and 68~per~cent confidence intervals for the mass-limited samples ($\log_{10} M_\star > 10$).}
    \label{tab:bootstrap_stats}
    \begin{tabular}{lcc}
        \hline
        \textbf{Sample} & \textbf{$\log_{10} \tilde{M}_\star$ [M$_\odot$]} & \textbf{Median $\log_{10}\mathrm{sSFR}$ [yr$^{-1}$]} \\
        \hline
        SDSS AGN hosts & $10.652^{+0.004}_{-0.003}$ & $-11.369^{+0.007}_{-0.005}$ \\[4pt]
        TNG100 centrals & $10.905^{+0.010}_{-0.013}$ & $-14.850^{+0.015}_{-0.014}$ \\[4pt]
        EAGLE centrals & $10.812^{+0.031}_{-0.015}$ & $-11.707^{+0.029}_{-0.039}$ \\
        \hline
    \end{tabular}
\end{table}

\section{sSFR distributions}
\label{sec:appendix_ssfr}

\begin{figure}
\centering
\includegraphics[width=\columnwidth]{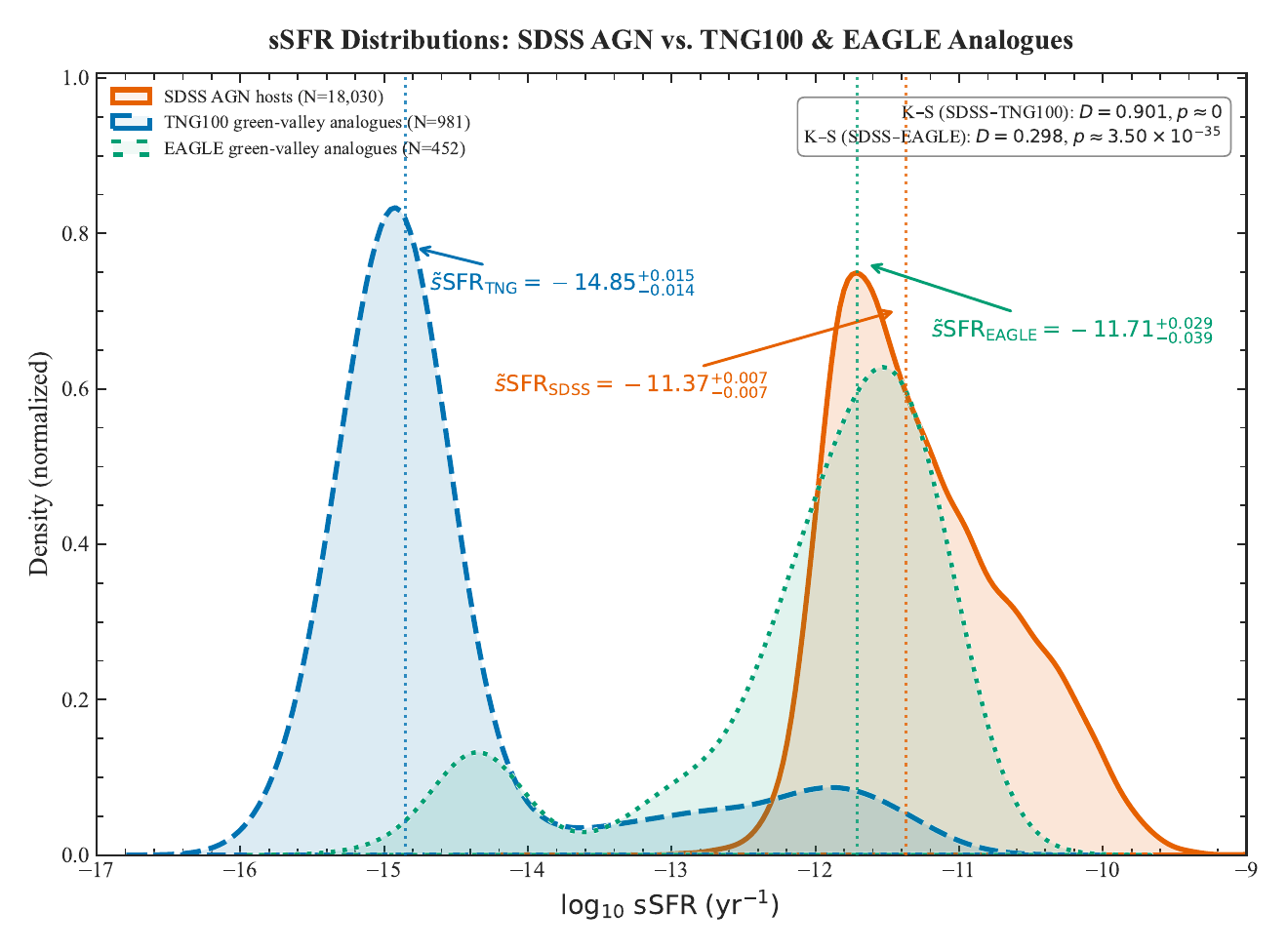}
\caption{Normalised KDEs of $\log_{10}\mathrm{sSFR}$ for massive ($\log_{10}M_\star>10$) SDSS AGN hosts (orange), TNG100 green-valley centrals (blue dashed), and EAGLE green-valley centrals (green dotted). Vertical dotted lines and arrows indicate bootstrapped median sSFRs with 68~per~cent confidence intervals. TNG is strongly over-quenched relative to SDSS, while EAGLE provides a much closer match to the observed distribution.}
\label{fig:ssfr_appendix}
\end{figure}

Fig.~\ref{fig:ssfr_appendix} presents the sSFR distributions for the mass-limited samples in a format complementary to Fig.~\ref{fig:cdfs}. TNG100 green-valley hosts are overwhelmingly quenched, with their sSFR distribution piled up at the imposed low-sSFR pile-up corresponding to formally zero instantaneous SFR (given the adopted $10^{-4}\,\mathrm{M_\odot\,yr^{-1}}$ SFR floor) and a median several dex below that of the SDSS AGN hosts. EAGLE predicts a broader, higher-sSFR distribution that is much closer to the SDSS AGN distribution and intermediate between SDSS and TNG100.

\bsp
\label{lastpage}
\end{document}